\documentclass[aps,prl,twocolumn,twoside,a4paper,10pt,amsmath,longbibliography,amssymb]{revtex4-1}
\usepackage{mathrsfs}
\usepackage{graphicx}
\usepackage{amsmath}
\usepackage{amssymb}
\usepackage{amsfonts}
\usepackage{color}
\usepackage{dcolumn}
\usepackage{bm}
\usepackage[dvipdfm, pdfstartview=FitH, CJKbookmarks=true, bookmarksnumbered=true, bookmarksopen=true, colorlinks=true, pdfborder=001, citecolor=blue, linkcolor=blue, linktocpage=true] {hyperref}

\def\ket#1{\left|#1\right\rangle}
\def\bra#1{\left\langle#1\right|}

\begin{document}

\title{Multi-particle Wannier states and Thouless pumping of interacting bosons }

\author{Yongguan Ke$^{1,2}$}
\author{Xizhou Qin$^{1}$}
\author{Yuri S. Kivshar$^{3}$}
\author{Chaohong Lee$^{1,2}$}
\altaffiliation{Email: lichaoh2@mail.sysu.edu.cn}

\affiliation{$^{1}$TianQin Research Center \& School of Physics and Astronomy, Sun Yat-Sen University (Zhuhai Campus), Zhuhai 519082, China}
\affiliation{$^{2}$State Key Laboratory of Optoelectronic Materials and Technologies, Sun Yat-Sen University (Guangzhou Campus), Guangzhou 510275, China}
\affiliation{$^{3}$Nonlinear Physics Center, Research School of Physics and Engineering, Australian National University, Canberra ACT 2601, Australia}

\date{\today}

\begin{abstract}
The study of topological effects in physics is a hot area, and only recently researchers were able to address the important issues of topological properties of interacting quantum systems. But it is still a great challenge to describe multi-particle and interaction effects. Here, we introduce multi-particle Wannier states for interacting systems with co-translational symmetry. We reveal how the shift of multi-particle Wannier state relates to the multi-particle Chern number, and study the two-boson Thouless pumping in an interacting Rice-Mele model. In addition to the bound-state Thouless pumping in which two bosons move unidirectionally as a whole, we find topologically resonant tunneling in which two bosons move unidirectionally, one by the other, provided the neighboring-well potential bias matches the interaction energy. Our work creates a new paradigm for multi-particle topological effects and lays a cornerstone for detecting interacting topological states.
\end{abstract}

\maketitle

For many decades, the physics has been dominated by the lattice symmetries and chemical composition, the key concepts in the classification and design of various materials.
However, it has recently been demonstrated that {\em topology} may be more important than symmetry in determining certain properties.
Topology is a subtle global property of the system governing how its parts connect.
Thouless pumping, the quantized transport in a slowly and cyclicly modulated periodic potential, is a typical {\em topological phenomenon}~\cite{Thouless1983}.
In single-particle systems, Thouless pumping connects the shift of Wannier state with the Chern number~\cite{King1993,Xiao2010}.
In addition to electronic systems, several atomic and photonic systems have been proposed to implement Thouless pumping~\cite{Wang2013,Mei2014,Wei2015,Ke2016a}.

Recently, Thouless pumpings of noninteracting cold atoms have been realized~\cite{Lohse2016,Nakajima2016,Lu2016}.
Since the atom-atom interactions can be tuned by Feshbach resonance,
cold atomic systems provide new opportunities to explore the interplay between topology and interaction~\cite{Xu2013,Zhou2015,Zeng2015,Zeng2016,Goldman2016,Tai2016,Lindner2017}.
Many-body polarization theory shows that the position shift of particles relates to the Berry phase of the ground state with a twisted angle~\cite{Ortiz1994,Xiao2010}.
However, it is still unknown how to generate an initial state for implementing multi-particle Thouless pumping.
The ongoing experiments urge us to develop an alternate framework of theory for interacting multi-particle systems.

An important step in Thouless pumping is how to prepare an initial state homogeneously populating a specific Bloch band.
Single-particle Wannier states can be given as unitary transformations of single-particle Bloch functions~\cite{Wannier1937}.
Because of the gauge dependence of Bloch functions, the single-particle Wannier states are strongly arbitrary.
To overcome this barrier, maximally localized Wannier states (MLWSs) have been proposed as a powerful tool for constructing localized orbits and a local probe for exploring electric polarization and orbital magnetization etc~\cite{Marzari1997,Marzari2012}.
Besides the electronic systems, MLWSs have been a versatile tool to construct lattice Hamiltonians for various periodic systems including  phonons~\cite{Rabe1995,Jorge2000,Giustino2006}, photons~\cite{Whittaker2003,Hermann2003,Takeda2006,Hartmann2008,Longhi2009}, and atoms~\cite{Jaksch1998,Bloch2008,Michele2012,Walters2013} etc.

It is a great challenge to extend the concept of Wannier state to multi-particle interacting systems.
As the interparticle interaction breaks the translational symmetry of individual particles, the multi-particle Wannier states cannot be constructed in a usual way~\cite{Koch2001, Hamann2009,Souza2000}.
Fortunately, although the interaction breaks the translational symmetry of individual particles, the particles as a whole may still have co-translational symmetry~\cite{Kolovsky2003,Fehske2008,Zhang2010}.
The corresponding eigenstates are identified as multi-particle Bloch states with center-of-mass (c.o.m) quaimomentum, which can be used to define the multi-particle Chern number~\cite{Qin2016a,Qin2016b}.
An important question naturally arises: Can we construct multi-particle Wannier states from multi-particle Bloch states?
If the answer is yes, can we establish the connection between the multi-particle Wannier state and the Chern number for a multi-particle Bloch band?

In this Letter, we construct the multi-particle Wannier states (MPWSs) as unitary transformation of multi-particle Bloch states.
The maximally localized multi-particle Wannier states (MLMPWSs) are generated by minimizing the spread functionals.
The MLMPWSs provide an orthogonal basis for constructing the effective lattice Hamiltonian for the isolated multi-particle Bloch band.
The reduced dimension of the effective Hamiltonian, which corresponds to the subspace for the isolated multi-particle Bloch band, will greatly benefit many-body calculations.
In the Thouless pumping, we find that the shift of MPWS is proportional to the Chern number of the filled multi-particle Bloch band.
We illustrate our formalism by studying the Thouless pumping of two interacting bosons in Rice-Mele lattices.

\emph{Theory}.---Let us consider an interacting $N$-particle system with $L$ $q$-site cells and assume the inter-particle interactions only depend on the relative distance between particles.
Here, we only consider bosons but the theory is also applicable to fermions when taking careful of boundary condition.
Imposing the period boundary condition, the system is invariant if all the particles are translated as a whole for integer cells.
The single-cell co-translational operator $\hat T_q$ can be defined as,
\begin{equation}
T_q |n_1,n_2,...,n_{qL}\rangle = |n_{q(L-1)+1},...,n_{qL},n_1,...,n_{q(L-1)}\rangle,
\end{equation} 
where $\left| \{n_j\}\right\rangle=|n_1,n_2,...,n_{qL}\rangle$ denotes the Fock state of $n_j$ particles in the $j$-th site.
The co-translational operator $T_q$ commutes with the Hamiltonian, $\hat T_q^{-1} \hat H \hat T_q=\hat H$.
Similar to the use of the shift operator~\cite{Kolovsky2003} and the single-particle translational operator~\cite{Fehske2008} in regular lattices, we alternately construct an orthogonal basis of the Hilbert space, $|\kappa,\textbf{n}\rangle={1\over{\sqrt{M}}}\sum_{j=0}^{M-1}{\exp(i\kappa qj)\hat T_q^j|\textbf{n}\rangle}$,
with the $j$-cell co-translational operator $\hat{T}_q^j=\left(\hat{T}_q\right)^j$.
Here, $|\textbf{n}\rangle$ is the seed state and $M (\le L)$ is the number of Fock states generated by repeatedly applying the co-translational operator on the seed state.
The set of Fock states $\{|\textbf{n}\rangle, \hat T_q|\textbf{n}\rangle,...,\hat T_q^{M-1}|\textbf{n}\rangle \}$ forms a translational cycle corresponding to the  c.o.m. quasi-momentum $\kappa=2\pi l/(qM)$ with $l=(0,2,...,M-1)$.
In the new basis, the Hamiltonian matrix can be block diagonalized as $\hat H = \oplus_{j=1}^L \hat H(\kappa_j)$,
where the matrix elements of $\hat H(\kappa_j)$ are given as $\langle \kappa_j,\textbf{n}'| \hat H |\kappa_j,\textbf{n}\rangle$.
One can give the multi-particle Bloch band by solving $\hat H(\kappa)|\psi_m(\kappa)\rangle=E_m(\kappa)|\psi_m(\kappa)\rangle$,
where the eigenstate $|\psi_m(\kappa)\rangle=\sum_{\textbf{n}}\psi_m(\kappa,\textbf{n})|\kappa,\textbf{n}\rangle$ is the multi-particle Bloch state in the $m$-th band with the c.o.m quasi-momentum $\kappa$ and the eigenvalue $E_m(\kappa)$~\cite{equalmomentum}.

Naturally, the MPWS for the $m$-th multi-particle Bloch band can be defined as
\begin{equation}
\left| {{W_m}(R)} \right\rangle  = \frac{1}{{\sqrt L }}\sum_{\kappa ,{\bf{n}}}^{} {{e^{ - i\kappa qR}}{\psi _m}(\kappa ,{\bf{n}})\left| {\kappa ,{\bf{n}}} \right\rangle },
\end{equation}
where $qR$ is the c.o.m. position of the MPWS.
The orthogonal relation,
$\left\langle {{{W_m}(R)}}
\mathrel{\left | {\vphantom {{{W_m}(R)} {{W_{m'}}(R')}}}
 \right. \kern-\nulldelimiterspace}
 {{{W_{m'}}(R')}} \right\rangle  = {\delta _{m,m'}}{\delta _{R,R'}}$,
make the MPWSs as a set of basis for exact theory.
The MPWSs are also arbitrary due to the freedom in choosing the phase of multi-particle Bloch states, $e^{i\theta(\kappa)}|\psi_m(\kappa)\rangle$.
Following the prescription for single-particle systems~\cite{Marzari1997,Marzari2012},
one can obtain unique MLMPWSs by minimizing the spread functional, $\Omega=\langle \hat{x}^2\rangle-\langle \hat{x}\rangle^2$ with the position operator $\hat x=N^{-1}\sum_{j=1}^{qL}{j \hat n_j}$ and the number operator $\hat n_j=\hat a_j^\dag \hat a_j$.
For an isolated multi-particle Bloch band, one can simply smooth the phases of MPWSs to obtain the MLMPWSs (see Supplementary Material).

Considering a slow cyclic driven system, $\hat H (\kappa,t)$ is invariant under the translation $\kappa\rightarrow\kappa+2\pi/q$ and $t\rightarrow t+T_B$, where $T_B$ is the driven period.
Similar to the undriven systems~\cite{Qin2016a,Qin2016b}, one can naturally define the multi-particle Chern number in the $(\kappa, t)$-plane
as,
\begin{equation}
  C_m = {\frac{1}{{2\pi }}{\int_0^{2\pi/q } {d \kappa \int_0^{{T_B}} {dt{{\cal F}_m}(\kappa ,t)} } }}.
  \label{Chern}
\end{equation}
where ${{\cal F}_m} = i \left( {\left\langle {{{\partial _t}{\psi _m}}}
 \mathrel{\left | {\vphantom {{{\partial _t}{\psi _m}} {{\partial _{\kappa} }{\psi _m}}}}
 \right. \kern-\nulldelimiterspace}
 {{{\partial _{\kappa} }{\psi _m}}} \right\rangle  - \left\langle {{{\partial _{\kappa} }{\psi _m}}}
 \mathrel{\left | {\vphantom {{{\partial _{\kappa} }{\psi _m}} {{\partial _t}{\psi _m}}}}
 \right. \kern-\nulldelimiterspace}
 {{{\partial _t}{\psi _m}}} \right\rangle } \right)$ is the Berry curvature for the eigenstates $\ket{\psi_m}$.
One may alternatively calculate the Chern number defined under the twisted boundary condition.
However, at the zero point of the twisted angle, because the eigenvalues are always the same at quasi-momenta $\kappa$ and $2\pi/q-\kappa$, there appear many degenerate points.
Therefore the Chern numbers defined with the twisted angle are unstable in the numerical calculations of our few-body systems.

If the initial state is the MPWS of the $m$-th band, the c.o.m. shift in one pumping cycle is given as
$\Delta P= \langle \hat{x}_m(T_B) \rangle - \langle  \hat{x}_m(0) \rangle$,
with $\langle  \hat{x}_m(t) \rangle = \langle W_m(R,t)| \hat x | W_m(R, t)\rangle$ being the c.o.m. position.
The relation between $\Delta P$ and the Chern number $C_m$ is given as (see Supplementary Material),
\begin{equation}
 \Delta P = q C_m.
 \label{DeltaP}
\end{equation}
This means that, through measuring the c.o.m. position shift, one can give the multi-particle Chern number.

\emph{Interacting Rice-Mele Model}.---The original Rice-Mele model describes non-interacting fermions in an asymmetric double-well superlattice~\cite{Rice1982}.
The asymmetric double-well optical superlattice can be created by superimposing a short-wavelength standing-wave $V_{s}(t)=-V_s\cos^2(\pi x/d )$ with period $d$ and a long-wavelength standing-wave $V_{l}(t)=-V_l\cos^2[\pi x/(2d)-\varphi(t)/2]$ with period $2d$, where $(V_{s},V_{l})$ denote the lattice depth~\cite{Wang2013,Nakajima2016}.
The relative phase $\varphi(t)=\varphi_0+\omega t$ can be controlled by a piezo-transducer-mounted mirror~\cite{Nakajima2016}, where $\omega$ is the modulation frequency.
Considering two repulsive ultracold Bose atoms in the above superlattice, the system obeys an interacting Rice-Mele model,
\begin{eqnarray}
  {\hat H} &=& \sum\limits_{j{\rm{ = }}1}^{2L} { \left[ (- J + (-1)^{j}\delta ) \hat a_{j}^\dag {\hat a_{j+1}}  + h.c.\right]} \nonumber \\
  &+&\sum\limits_{j=1}^{2L} {\left[(-1)^{j} \Delta  {\hat n_j}+\frac{U}{2} {\hat n_{j}}({\hat n_{j}} - 1)\right]},
  \label{RiceMele}
\end{eqnarray}
with the site index $j$, the bosonic creation (annihilation) operators $\hat a_j^\dag$ ($\hat a_j$) for the $j$-th site, and the total number of lattice sites $2L$.
Here, the interaction strength $U$ can be tuned by the Feshbach resonance~\cite{Bloch2008}, $- J + (-1)^{j}\delta$  stands for the hopping strength between the $j$-th and $(j+1)$-th sites, and the on-site energy is either $-\Delta$ for the odd sites or $+\Delta$ for the even sites.
In the Thouless pumping, $\delta$ and $\Delta$ are tuned according to $\delta=\delta_0 \sin(\omega t+\varphi_0)$ and $\Delta=\Delta_0 \cos( \omega t+\varphi_0)$, respectively.
Here, the pumping cycle is given as $T_B=2\pi/\omega$, $\delta_0$ is hopping modulation strength and $\Delta_0$ is the on-site modulation strength (see Supplementary Material).

%%%%%%%%%%%%%%%%%%%%
\begin{figure}[!htp]
\begin{center}
\includegraphics[width=1\columnwidth]{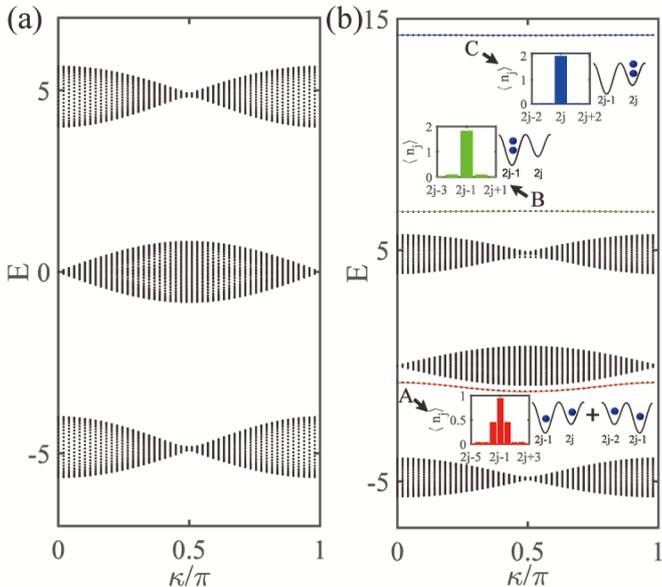}
\end{center}
\caption{ Multi-particle Bloch bands and MLMPWSs.
(a) Multi-particle Bloch bands of two non-interacting bosons.
The parameters are chosen as $J=1,~\delta_0=0.5,~\Delta_0=2, \omega t+\varphi_0=0,~U=0$ and the system size is $2L=98$.
(b) Multi-particle Bloch bands of two interacting bosons.
Three isolated multi-particle Bloch bands (A, B, C) are separated from the continuum scattering-state bands.
Insets (bar graphs): the density distributions of MLMPWSs for the isolated bands (A, B, C).
The corresponding schematic pictures for the MLMPWSs are shown in the right hand side.
The parameters are the same as those in (a) except for $U=10$.}
\label{WannierState}
\end{figure}
%%%%%%%%%%%%%%%%%%%%

\emph{Multi-particle Bloch Bands and MLMPWSs}.---In non-interacting systems, the two bosons hop independently in the Rice-Mele lattices and there are only continuum bands for scattering states.
If the double-well bias is sufficiently large, there are three kinds of scattering states: (i) both two bosons in the lower sublattices, (ii) one boson in the lower sublattices and the other in the higher sublattices and (iii) both two bosons in the higher sublattices.
Fig.~\ref{WannierState}(a) shows the multi-particle Bloch bands for two non-interacting bosons.
The parameters are chosen as $J=1,~\delta_0=0.5,~\Delta_0=2,~\omega t+\varphi_0=0,~U=0$ and the system size is $2L=98$.
The three continuum bands from the bottom to the top are corresponding to type-(i), type-(ii) and type-(iii) scattering states, respectively.

In interacting systems, in addition to the continuum bands, there appear three isolated bands (denoted by A, B and C), see Fig.~\ref{WannierState}(b).
All of the parameters are the same as those in Fig.~\ref{WannierState}(a) except for $U=10$.
To understand the isolated bands, we respectively calculate the density distributions of MLMPWSs corresponding to the isolated bands (A, B, C), see the (red, green, blue) bar graphs in Fig.~\ref{WannierState}(b), where the center cell is chosen as the $j$-th cell.
The particle density distribution of the MLMPWS is defined as $\langle n_j\rangle=\langle W_m(R)|\hat n_j |W_m(R)\rangle$, which exponentially decays away from the center cell and resembles the single-particle Wannier states.
For the isolated bands B and C, their MLMPWSs have the two bosons almost respectively staying in the same odd and even sites, see the insets by the right hand side of green  and blue bar graphs in Fig.~\ref{WannierState}(b).
These MLMPWSs are naturally identified as bound states, whose effective Hamiltonian is given below via degenerate perturbation theory.
However, the MLMPWSs for the isolated band A can not be explained by degenerate perturbation theory, due to their energy scale is quite near the type-(ii) scattering states.
For the isolated band A, its MLMPWSs have the two bosons approximate in the superposition states located in two neighboring double-well cells, see the inset by the right hand side of red bar graph in Fig.~\ref{WannierState}(b).

By using the  MLMPWSs, one can derive the tight-binding Hamiltonians for all three isolated bands.
The hopping energy is given as $J_b=\langle W_m(R)|\hat H |W_m(R+1)\rangle$ and the on-site energy is given as $\epsilon_b=\langle W_m(R)|\hat H |W_m(R)\rangle$.
In the quasi-momentum space, the energy bands for the effective Hamiltonians are formulated as $E(\kappa)=2J_b\cos(\kappa q)+\epsilon_b$, see the (red, green, blue) solid lines covering the (A, B, C) isolated bands, respectively.
This means that the single-band tight-binding Hamiltonians derived from the MLMPWSs can well describe the behavior of the isolated bands.

%%%%%%%%%%%%%%%%%%%%%
\begin{figure*}[!htp]
\begin{center}
\includegraphics[width=2\columnwidth]{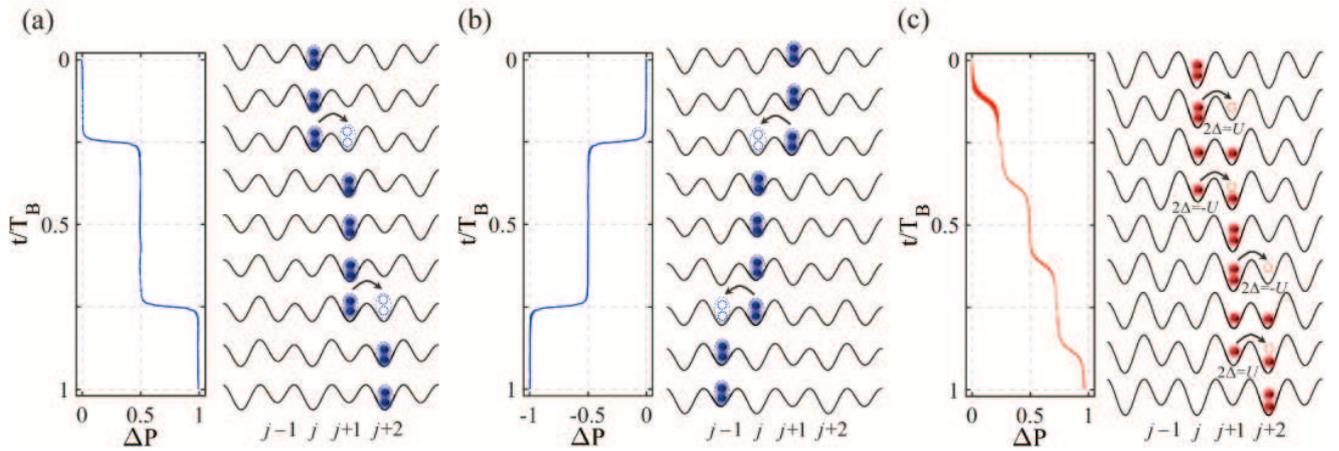}
\end{center}
\caption{ Thouless pumping of interacting particles.
Mean position shift as a function of time for the initial state in the lower well (a) and upper well (b) in Thouless pumping of bound states. The schematic diagram of (a) and (b) show that the two bosons initially prepared in either the lower or upper well will be unidirectionaly transported a whole, respectively. The parameters are set as $J=1,~\delta_0=0.8,~\Delta_0=2,~U=30,~\varphi_0=0$ and $\omega=0.005$.
(c) Mean position shift as a function of time in topologically resonant tunneling. The schematic diagram of (c) show that the two bosons initially prepared in the lower well will be unidirectionaly transported through the double-well barrier one by one when the bias matches the interaction.The parameters are set as $J=1,~\delta_0=0.8,~\Delta_0=20,~U=30,~\varphi_0=0$ and $\omega=0.005$.}
\label{systematic}
\end{figure*}
%%%%%%%%%%%%%%%%%%%%%

\emph{Thouless pumping of bound states}.---Below we discuss the Thouless pumping associated with the isolated bands B and C.
If $U \gg (J,~\delta,~\Delta)$, the two bosons staying in the same site will form a bound state~\cite{Winkler2006,Valiente2008,Qin2014,Fukuhara2013,Preiss1229}.
Applying the degenerate perturbation theory up to the second order~\cite{Takahashi1977,Qin2014}, the effective Hamiltonian for the bound states is given as (see Supplementary Material),
\begin{equation}
  \Hat{H}_{eff}=\sum\limits_{j=1}^{2L} {  J_{eff}(j) \hat b_j^\dag {\hat b_{j+1}} + h.c. } +\sum\limits_{j=1}^{2L}2(-1)^{j}\Delta \hat b_j^\dag {\hat b_j},
\label{HamEff}
\end{equation}
Here, $\hat b_j^{\dag} \ket 0 = \ket {1}_j^{b} \Leftrightarrow \ket {2}_j$ which is short for $|0,...,2_j,...0\rangle$.
The effective hopping energy between $j$ and $j+1$-th sites is $J_{eff}(j) = 2(J +(-1)^{j+1} \delta)^2/U$ and the uniform on-site energy shift $U + 2{(J + \delta )^2}/U + 2{(J - \delta )^2}/U$  is neglected.
The bound state can be viewed as a single quasi-particle hopping in the double-well superlattices.
The Chern numbers of the two bound-state bands are +1 and -1, consistent with the multi-particle Chern numbers given by Eq.~\eqref{Chern} (see Supplementary Material).

We simulate the Thouless pumping of bound states in a 58-site system.
In our simulation, the parameters are set as $J=1,~\delta_0=0.8,~\Delta_0=2,~U=30,~\varphi_0=0$ and $\omega=0.005$.
We choose two initial states of two bosons respectively staying in the $29$ and $30$-th sites, which respectively occupy the isolated bands B and C.
These initial states can be prepared via the current experiment technique for two-boson quatum walks~\cite{Preiss1229}.
Both two initial states have over $99.4\%$ projection on the MLMPWSs for the corresponding isolated bands at time $t=0$.
Fig.~\ref{systematic}(a)-(b) shows the c.o.m. position shifts $\Delta P$ as a function of time for the initial states $|2\rangle_{29}$ and $|2\rangle_{30}$, respectively.
The c.o.m. position shifts $0.993$ unit cell to the right for the initial state $|2\rangle_{29}$ and $0.995$ unit cell to the left for the initial state $|2\rangle_{30}$ in one pumping cycle.
Both of them are very close to the corresponding multi-particle Chern numbers given by Eq.~\eqref{Chern} .
During the pumping process, the time-evolution of MLMPWSs show the bound states are unidirectionally transported as a whole, see the schematic diagrams in Fig.~\ref{systematic}(a)-(b).

\emph{Topologically resonant tunnelings}.---Below we discuss the Thouless pumping associated with the isolated band A.
If $U \gg (J,~\delta)$, the bound state breaks down and resonant tunneling happens between $\ket {2}_j$ and $\ket {1}_j\ket {1}_{j+1}$ or $\ket {1}_{j-1}\ket {1}_j$ when $U$ is comparable to $2\Delta$~\cite{Lee2008,Cheinet2008,Xinfang2015}.
Here, $|1\rangle_{j}|1\rangle_{k}$ is short for $|0,...,1_{j},...,1_{k},...,0\rangle$.
However, the interplay between the band topology and the resonant tunneling will unidirectionally transport the two bosons one by one for integer cells in one pumping cycle and so that we call it the topologically resonant tunneling, see the schematic diagram in Fig.~\ref{systematic}(c).
In our simulation, the initial state is chosen as $|2\rangle_{29}$ (both two bosons in the $29$-th site) for a 58-site system, and the parameters are set as $J=1,~\delta_0=0.8,~\Delta_0=20,~U=30,~\varphi_0=0$ and $\omega=0.005$.
The initial state have $96.4\%$ overlap with the corresponding MLMPWS at time $t=0$.
The corresponding multi-particle Chern number is +1 (see Supplementary Material).
In Fig.~\ref{systematic}(c), we show the c.o.m. position shift as a function of time.
The c.o.m. position is shifted $0.968$ unit cell to the right in one pumping cycle.
The slight derivation from the Chern number of the isolated band is caused by the imperfect initial state.

The topological resonant tunneling is a result of the interplay among the inter-particle interaction, the double-well bias and the band topology.
To make it more clear, we analyse a non-interacting system of $\varphi_0=\pi/2$ and the initial state given as the MLMPWS $\hat a_{29}^{\dag}\hat a_{30}^{\dag}|0\rangle$.
Such an initial state can be decomposed as $\hat a_{29}^{\dag}\hat a_{30}^{\dag}|0\rangle=\frac{1}{\sqrt{2}}((\hat a_S^{\dag})^2-(\hat a_A^{\dag})^2)|0\rangle$,
with the symmetric single-particle state $\hat a_S^{\dag}|0\rangle=\frac{1}{\sqrt{2}}(\hat a_{29}^{\dag}+\hat a_{30}^{\dag})|0\rangle$ and the anti-symmetric single-particle state
$\hat a_A^{\dag}|0\rangle=\frac{1}{\sqrt{2}}(\hat a_{29}^{\dag}-\hat a_{30}^{\dag})|0\rangle$~\cite{Tai2016}.
The symmetric (anti-symmetric) states respectively fill the lower (upper) bands of the single-particle Rice-Mele model.
Thouless pumping from the symmetric (anti-symmetric) states will make the c.o.m. position shift one unit cell along right (left) in one pumping cycle~\cite{Lohse2016}.
Because the initial state $\hat a_{29}^{\dag}\hat a_{30}^{\dag}|0\rangle$ has equal probability of two bosons in the same symmetric and anti-symmetric single-particle states, there is no c.o.m. transport.

\emph{Summary and Discussions}.---We have put forward a new concept of MPWSs and studied the Thouless pumping of strongly interacting bosons in Rice-Mele lattices.
The set of MPWSs provides an orthogonal basis and their c.o.m. shifts in Thouless pumping are proportional to the Chern numbers of the filled multi-particle Bloch bands.
By minimizing the spread functional of MPWSs, one can generate the MLMPWSs, which can used as perfectly initial states for implementing Thouless pumping.
If the interaction energy is much larger than the double-well bias, a bound state will be unidirectionally transported through one double-well cell as a whole during one pumping cycle.
However, if the double-well bias may balance the interaction, the interplay between resonant tunneling and band topology will drive the two bosons unidirectionally through the barrier one by one.
In both Thouless pumping of bound state and topologically resonant tunneling, the two bosons will be shifted integer double-well cells in one pumping cycle.

To apply the MPWS concept to many-body problems, one needs to find an efficient method to reduce the computational resource.
Matrix product state is a powerful method for many-body calculations~\cite{Perez2007,Schollw0ck2011}, such as the many-body dispersion relations~\cite{Pirvu2011,Haegeman2012,Pirvu2012}.
MPWS may be expressed with the matrix product sate representation and then one can used it for exploring fractional topological states.

Note added: In the process of preparing our manuscript, we became aware of the Thouless pumping of three attractive photons in one dimensional nonlinear resonator arrays~\cite{Tangpanitanon2016}.

We acknowledge J. Huang, S. Wu, Q. Ye and H. Zhong for discussions.
This work was supported by the National Natural Science Foundation of China (NNSFC) under Grants No. 11374375 and No. 11574405, and the Australian Research Council (ARC).

%\bibliography{ResonantBound}
%

%\end{document} 

%\begin{document}

\newpage

\section*{Supplementary Material}

In this Supplementary Material, we provide more details about the maximally localized multi-particle Wannier state, the relation between the multi-particle Wannier state and the multi-particle Chern number, the derivation of the interacting Rice-Mele Hamiltonian and the topological properties of multi-particle Bloch bands.  

\subsection*{Maximally localized multi-particle Wannier states} \label{MLMPWS}

A general approach to obtain maximally localized multi-particle Wannier states (MLMPWSs) is to minimize the spread functional of the multi-particle Wannier states (MPWSs) in the $\mathcal{M}$ cluster band~\cite{Marzari1997,Marzari2012}:
\begin{equation}
  \Omega=\sum\limits_{m\in\mathcal{M}} \langle x^2\rangle_m-\langle x\rangle_m^2,
\end{equation}
with $\langle x^2\rangle_m=\langle W_m(0)|\hat x^2|W_m(0)\rangle$ and $\langle x\rangle_m=\langle W_m(0)|\hat x|W_m(0)\rangle$.
The spread functional can be decomposed as,
\begin{equation}
  \Omega=\Omega_I+\Omega_D + \Omega_{OD},
\end{equation}
with
\begin{eqnarray}
 && \Omega_I =\sum\limits_{m \in\mathcal{M}}  \sum\limits_{m' \notin\mathcal{M},R} {{{\left| {\left\langle {{W_{m'}}(R)|\hat x|{W_m}(0)} \right\rangle } \right|}^2}}
  \label{Ispread}, \\
 &&{\Omega _D} = \sum\limits_{m\in\mathcal{M}} {\sum\limits_{R \ne 0}^{} {{{\left| {\left\langle {{W_m}(R)|\hat x|{W_m}(0)} \right\rangle } \right|}^2}} },
  \label{Dspread} \\
 && {\Omega _{OD}} = \sum\limits_{m,m'\in\mathcal{M}: m \ne m'}^{} {\sum\limits_R^{} {{{\left| {\left\langle {{W_{m'}}(R)|\hat x| {W_m}(0)}\right\rangle } \right|}^2}} }.
 \label{ODspread}
\end{eqnarray}

Below we will transfer the expressions to the momentum space.
Applying the position operator on the multi-particle Wannier state (MPWS), one can obtain,
\begin{equation}
  \hat x\left| {{W_m}(0)} \right\rangle   = \frac{1}{L}\sum\limits_{\kappa ,{\bf{n}},j}^{} {{e^{i\kappa qj}}{\psi _m}(\kappa ,{\bf{n}})qjT_q^j\left| {\bf{n}} \right\rangle }.
\end{equation}
Here, because of the periodic boundary condition, the mean positions of the seed states are set to be 0.
In the limits of large $L$, one can replace the summation over quasi-momentum $\kappa$ by the form of continuous integral, that is,
\begin{eqnarray}
 && \hat x\left| {{W_m}(0)} \right\rangle = \frac{q}{{2\pi }}\sum\limits_{{\bf{n}},j}^{} {\int_0^{2\pi /q} {{e^{i\kappa qj}}{\psi _m}(\kappa ,{\bf{n}})qjd\kappa } T_q^j\left| {\bf{n}} \right\rangle } \nonumber \\
 &=&\frac{q}{{2\pi }}\sum\limits_{{\bf{n}},j}^{} {\int_0^{2\pi /q} {{e^{i\kappa qj}}i\frac{\partial }{{\partial \kappa }}{\psi _m}(\kappa ,{\bf{n}})d\kappa } T_q^j\left| {\bf{n}} \right\rangle } \nonumber \\
 &-&\frac{q}{{2\pi }}\sum\limits_{{\bf{n}},j}^{} {\int_0^{2\pi /q} {i\frac{\partial }{{\partial \kappa }}\left({e^{i\kappa qj}}{\psi _m}(\kappa ,{\bf{n}})\right)d\kappa } T_q^j\left| {\bf{n}} \right\rangle },
\end{eqnarray}
Because ${e^{i\kappa qj}}{\psi _m}(\kappa ,{\bf{n}})$ is a periodic function with periodicity $2\pi/q$, the integral of $\frac{\partial }{{\partial \kappa }}\left({e^{i\kappa qj}}{\psi _m}(\kappa ,{\bf{n}})\right)$ over one period is 0.
Thus only the first term is preserved in the above equation.
Then the elements $\left\langle {{W_{m'}}(R)} \right|\hat x\left| {{W_m}(0)} \right\rangle$ in equations~\eqref{Ispread}-\eqref{ODspread} can be expressed as,
\begin{eqnarray}
&&\left\langle {{W_{m'}}(R)} \right|\hat x\left| {{W_m}(0)} \right\rangle \nonumber \\
&=& \frac{q}{{2\pi }}\int_0^{2\pi /q} {e^{i\kappa q R} \left\langle {{\psi _{m'}}(\kappa )} \right|i\frac{\partial }{{\partial \kappa }}\left| {{\psi _m}(\kappa )} \right\rangle d\kappa },
\label{SP}
\end{eqnarray}
where we have used $\left\langle {{\bf{n'}}} \right|T_q^{ - j'}T_q^j\left| {\bf{n}} \right\rangle =\delta_{j,j'}\delta_{\bf{n},\bf{n}'}$.
Substituting  equation~\eqref{SP} into equations~\eqref{Ispread}-\eqref{ODspread}, one can obtain~\cite{Michele2012},
\begin{eqnarray}
  &&  \Omega_{D}=\sum\limits_{m \in\mathcal{M}}{ \langle \left| A_{m,m}(\kappa)-\langle A_{m,m}(\kappa)\rangle_{\mathcal{B}}\right|^2\rangle_{\mathcal{B}}},\\
  &&  \Omega_{OD}=\sum\limits_{m \in\mathcal{M}}  \sum\limits_{m' \in\mathcal{M},m'\ne m} {\langle \left|A_{m',m}(\kappa)\right|^2\rangle_{\mathcal{B}}}, \\
  && \Omega_I=\sum\limits_{m \in\mathcal{M}}  \sum\limits_{m' \notin\mathcal{M}} {\langle \left|A_{m',m}(\kappa)\right|^2\rangle_{\mathcal{B}}},
\end{eqnarray}
where $A_{m',m}(\kappa) = \left\langle {{\psi _{m'}}(\kappa )} \right|i\frac{\partial }{{\partial \kappa }}\left| {{\psi _m}(\kappa )} \right\rangle$ and $\langle y(\kappa) \rangle_{\mathcal{B}}=\frac{q}{2\pi}\int_{0}^{2\pi/q}{y(\kappa) d\kappa}$.
One can prove that $\Omega_{I}$ is invariant under the unitary transformation $|\psi_m(\kappa)\rangle \rightarrow \sum_{n\in\mathcal{M}} U_{m,n}(\kappa)|\psi_n(\kappa)\rangle$, which mixes the different bands.

To minimize the spread functional, one has to find the optimal unitary transformation to minimize $\Omega_D$ and $\Omega_{OD}$.
For the isolated band in one dimension, $\Omega_{OD}$ is not present and the problem is reduced to minimize $\Omega_D$.
It is straightforward to make~\cite{Marzari1997,Marzari2012}
\begin{equation}
   A_{m,m}(\kappa)=\langle A_{m,m}(\kappa)\rangle_{\mathcal{B}}.
   \label{Smooth}
\end{equation}
One can give a computational expression for $\Omega_D$ in the form of discrete mesh in momentum space.
The left hand side of the equation is given as
\begin{equation}
  A_{m,m}(\kappa_j)=-\frac{1}{d\kappa}{\rm Im} \ln\langle \psi_m(\kappa_j)|\psi_m(\kappa_{j+1})\rangle,
\end{equation}
and the right hand side of the equation is given as
\begin{equation}
 \langle A_{m,m}(\kappa_j)\rangle_{\mathcal{B}}=-\frac{1}{d\kappa L} {\rm Im} \ln\prod_{j=1}^L\langle \psi_m(\kappa_j)|\psi_m(\kappa_{j+1})\rangle.
\end{equation}
Here, $L$ is the number of $\kappa$ in the first Brillouin zone, $d\kappa=2\pi/(qL)$, and the last term $\langle \psi_m(\kappa_L)|\psi_m(\kappa_{L+1})\rangle=\langle \psi_m(\kappa_L)|\psi_m(\kappa_1)\rangle$ due to the periodicity of $|\psi_m(\kappa)\rangle$ in momentum space.
To realize equation~\eqref{Smooth}, one can make a gauge transformation of the $|\psi_m(\kappa_j)\rangle$ as
\begin{equation}
  |\psi_m(\kappa_j)\rangle=V_j|\psi_m(\kappa_j)\rangle,
\end{equation}
where
\begin{eqnarray}
 V_1&=&1, \nonumber \\
 V_{j+1}&=&e^{-i\theta_j} e^{i\theta_{ave}}V_j.
\end{eqnarray}
Here, $\theta_j={\rm Im} \ln\langle \psi_m(\kappa_j)|\psi_m(\kappa_{j+1})\rangle$ and $\theta_{ave}=\frac{1}{L}{\rm Im} \ln \prod_{j=1}^L \langle \psi_m(\kappa_j)|\psi_m(\kappa_{j+1})\rangle$.

\subsection*{Relation between the MPWS and the multi-particle Chern number} \label{relation}

From equation~\eqref{SP} and the translational symmetry of MPWSs, the c.o.m. position of the MPWS for the $m$-th band at time $t$ is given as
\begin{eqnarray}
 && \langle  x_m(t) \rangle= \left\langle {{W_{m}}(R,t)} \right|\hat x\left| {{W_m}(R,t)} \right\rangle \nonumber \\
  &=& qR+ \frac{q}{{2\pi }}\int_0^{2\pi /q} {\left\langle {{\psi _{m}}(\kappa,t )} \right|i\frac{\partial }{{\partial \kappa }}\left| {{\psi _m}(\kappa,t )} \right\rangle d\kappa }.
\end{eqnarray}
We define the c.o.m. shift from time $t$ and $t+dt$  as
\begin{equation}
 \partial P= \langle  x_m(t+dt) \rangle-\langle  x_m(t) \rangle.
\end{equation}
Because $\langle x_m(t)\rangle$ is a continuous function, the c.o.m. shift in one pumping cycle $T_B$ is given as
  \begin{eqnarray}
  && \Delta P = \int_0^{T_B}\frac{\partial P}{\partial t}dt \nonumber \\
  &=& \frac{q}{{2\pi }}\int_0^{T_B}\int_0^{2\pi /q} {\frac{\partial}{\partial t}\left(\left\langle {{\psi _{m}}(\kappa,t )} \right|i\frac{\partial }{{\partial \kappa }}\left| {{\psi _m}(\kappa,t )} \right\rangle \right)d\kappa dt} \nonumber \\
  & =&q C_m,
\end{eqnarray}
with the multi-particle Chern number
$C_m = {\frac{1}{{2\pi }}{\int_0^{2\pi/q } {d \kappa \int_0^{{T_B}} {dt{{\cal F}_m}(\kappa ,t)} } }}$
and the Berry curvature
${{\cal F}_m} = i \left( {\left\langle {{{\partial _t}{\psi _m}}}
 \mathrel{\left | {\vphantom {{{\partial _t}{\psi _m}} {{\partial _{\kappa} }{\psi _m}}}}
 \right. \kern-\nulldelimiterspace}
 {{{\partial _{\kappa} }{\psi _m}}} \right\rangle  - \left\langle {{{\partial _{\kappa} }{\psi _m}}}
 \mathrel{\left | {\vphantom {{{\partial _{\kappa} }{\psi _m}} {{\partial _t}{\psi _m}}}}
 \right. \kern-\nulldelimiterspace}
 {{{\partial _t}{\psi _m}}} \right\rangle } \right)$.

\subsection*{Derivation of the interacting Rice-Mele Hamiltonian} \label{secondQ}

We show how to derive the interacting Rice-Mele Hamiltonian~(8) in the main text.
The motion of interacting bosons in the superlattices is governed by
\begin{eqnarray}
&&\hat H =\int \hat \psi^\dag(x)\hat H_0 \hat \psi(x) dx +\frac{g}{2} \int \hat \psi^\dag(x)\hat \psi^\dag(x) \hat \psi(x) \hat \psi(x) dx, \nonumber \\
&&  \hat H_0 = \frac{{\hat p_x^2}}{{2m}} - {V_s}{\cos ^2}\left( {\frac{\pi }{d}x} \right) - {V_l}{\cos ^2}\left( {\frac{\pi }{{2d}}x - \varphi /2} \right). \nonumber \\
\end{eqnarray}
Here, the field operator $\hat \psi^{\dag} (x)$ ($\hat \psi (x)$) creates (destroys) a boson at position $x$, $g=\frac{4\pi \hbar^2}{M} a$ denotes the interaction factor with the scattering length $a$ and $\hat p_x$ is the particle momentum.
We make a transformation $x'=\pi x/d$ and express the Hamiltonian in terms of $x'$,
\begin{equation}
  \hat H_{0}' =  - \frac{{{\partial ^2}}}{{{\partial}x'^2}} - \frac{{{V_s}}}{2}\cos \left( {2 x'} \right) - \frac{{{V_l}}}{2}\cos \left( {{x'} - \varphi } \right),
  \label{origin}
\end{equation}
where $V_s$ and $V_l$ are in the units of $E_r=\frac{\pi^2\hbar^2}{2md^2}$.

The Wannier states ${|w_j\rangle}$ of the lowest band of $ - \frac{{{\partial ^2}}}{{{\partial}x'^2}} - \frac{{{V_s}}}{2}\cos \left( {2x'} \right)$ form an orthogonal basis and $w_j(x')$ is localized at the $j$-th lattice when the lattice is deep enough.
One can expand the Hamiltonian~\eqref{origin} in term of the Wannier basis $\{|w_j\rangle\}$,
\begin{equation}
  {\hat H'} =  \sum\limits_{<i,j>}^{} {\left\langle {{w_{i}}} \right|\hat H_{0}' \left| {{w_j}} \right\rangle \hat a_i^\dag \hat a_j}+\frac{U}{2}\sum\limits_{j}\hat n_j(\hat n_j-1),
  \label{TB}
\end{equation}
where the interaction strength $U=g \int |w_j(x')|^4 dx'$ is in the unit of $E_r$.
As the lattice is sufficiently deep, one can only consider the contribution of the nearest neighboring hopping and on-site energy to the matrix element $\left\langle {{w_{i}}} \right|\hat H_{0}' \left| {{w_j}} \right\rangle$.
For the on-site energy,
\begin{eqnarray}
&&\left\langle {{w_{j}}} \right|\hat H_{0}' \left| {{w_j}} \right\rangle = -\frac{V_l}{2}\left\langle {{w_j}} \right|\cos \left( {{x'} - \varphi } \right)\left| {{w_j}} \right\rangle \nonumber  \\
&=& -\frac{V_l}{2}(-1)^j \cos ( \varphi ) \left\langle {{w_{0}(x')}} \right| \cos \left( {x'} \right)  \left| {{w_0(x')}} \right\rangle, \nonumber
\end{eqnarray}
where we have neglected the energy constant $ \langle w_j|- \frac{{{\partial ^2}}}{{{\partial}x'^2}} - \frac{{{V_s}}}{2}\cos \left( {2x'} \right)|w_j \rangle$.
For the nearest neighboring hopping energy,
\begin{eqnarray}
&&\left\langle {{w_{j}}} \right|\hat H_{0}' \left| {{w_{j+1}}} \right\rangle \nonumber \\
&=& -J - \frac{V_l}{2}\left\langle {{w_{j}}} \right|\cos \left( {x' - \varphi } \right) \left| {{w_{j+1}}} \right\rangle \nonumber \\
&=& -J -\frac{V_l}{2} {( - 1)^j}\sin \left( \varphi  \right)  \langle {{w_{0}(x'+\frac{\pi}{2})}} | \cos \left( {x' } \right)  | {{w_0(x'-\frac{\pi}{2})}} \rangle \nonumber
\end{eqnarray}
where $J=\left\langle {{w_{j}}} \right| \frac{{{\partial ^2}}}{{{\partial}x'^2}} + \frac{{{V_s}}}{2}\cos \left( {2x'} \right) \left| {{w_{j+1}}} \right\rangle$.
Then Hamiltonian~\eqref{TB} is derived as Hamiltonian~(7) by making
$\delta=\delta_0 \sin(\varphi)$ and $\Delta=\Delta_0 \cos(\varphi)$,
where
\begin{eqnarray}
 && \delta _0} =  - \frac{{{V_l}}}{2}\int {w_0(x' + \frac{\pi}{2} )\cos \left( {x'} \right)w_0(x' - \frac{\pi}{2} )dx', \nonumber \\
 && {\Delta _0} =  - \frac{{{V_l}}}{2}\int {w_0(x')\cos \left( {x'} \right)w_0(x')dx'}.
\end{eqnarray}

\subsection*{Multi-particle Bloch bands and their Chern numbers}

\subsubsection*{Topological bound states} \label{MPBB1}

When $U \gg (J,~\delta,~\Delta)$, one can treat non-interacting Hamiltonian $\hat{H}_0$ as a perturbation to the interaction term $\hat{H}_{int}$.
The interaction term $\hat{H}_{int}$ has two eigenvalues, $E_j=U$ with degenerate eigenstates $\ket{2}_j$  forming a subspace $\mathcal{U}$,
and $E_{j,k} =0$ with degenerate eigenstates $\ket{1}_j\ket{1}_k$ forming the orthogonal complement subspace $\mathcal{V}$.
We respectively define the projection operators upon $\mathcal{U}$ and $\mathcal{V}$ as
\begin{eqnarray}
&\hat P&=\sum\limits_j{\ket{2}_j\bra{2}_j}, \nonumber \\
&\hat S&=\sum\limits_{j\ne k}{{1\over {E_j-E_{j,k}}}\ket{1}_j\ket{1}_k\bra{1}_k\bra{1}_j}.
\end{eqnarray}
Applying the degenerate perturbation theory up to the second order~\cite{Takahashi1977,Qin2014}, the effective Hamiltonian for the subspace $\mathcal{U}$ is given as
\begin{equation}
  \Hat{H}_{eff}=E_j\hat{P}+\hat{P} \hat{H}_{0} \hat{P}+\hat{P}\hat{H}_0\hat{S}\hat{H}_0\hat{P}.
\end{equation}
After careful calculation, one can obtain the effective Hamiltonian for the bound states,
\begin{eqnarray}
&  \Hat{H}_{eff}=\sum\limits_{j=1}^{2L} {  \frac{2(J +(-1)^{j+1} \delta)^2}{U} \hat b_j^\dag {\hat b_{j+1}} + h.c. } \nonumber \\
&  +\sum\limits_{j=1}^{2L}2(-1)^{j}\Delta \hat b_j^\dag {\hat b_j},
\label{HamEff}
\end{eqnarray}
where $\hat b_j^{\dag} \ket 0 =\ket {1}_j^b \Leftrightarrow \ket {2}_j$ and the uniform on-site energy shift $U + 2{(J + \delta )^2}/U + 2{(J - \delta )^2}/U$  is neglected.

%%%%%%%%%%%%%%%
\begin{figure}[!htp]
\begin{center}
\includegraphics[width=1\columnwidth]{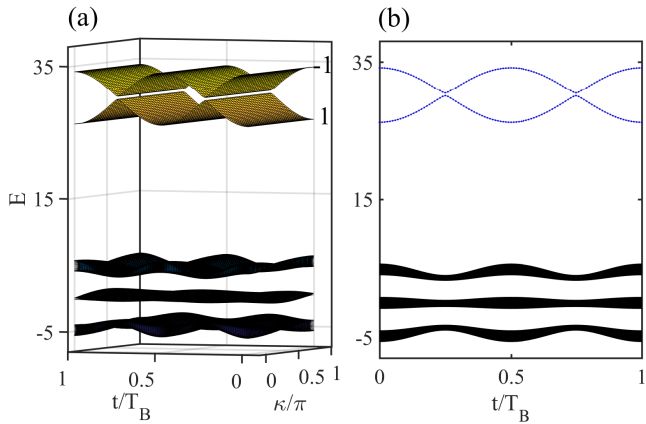}
\end{center}
\caption{Multi-particle Bloch band. (a) The 3D view of Bloch band in the $\kappa-t$ Brillouin zone. (b) The $t-E$ view of Bloch band. The parameters are chosen as $J=1,~\delta_0=0.8,~\Delta_0=2,~\varphi_0=0, ~U=30$ and the system size is 58.}
\label{Spectrum1}
\end{figure}
%%%%%%%%%%%%%%%

In Fig.~\ref{Spectrum1} (a) and (b), we show the multi-particle Bloch band of the Hamiltonian~(7).
The parameters are set as $J=1,~\delta_0=0.8,~\Delta_0=2, ~\varphi_0=0, ~U=30$ and the system size is set as $2L=58$.
There are $59$ bands: the highest two bands corresponding to bound states
and the other bands corresponding to scattering states.
We can confirm the bound-state bands by calculating the energy spectrum of the effective Hamiltonian~\eqref{HamEff} with the same parameters, see the blue dots covering the energy bands of bound states in Fig.~\ref{Spectrum1} (b).
The continuum scattering-state bands include three cluster bands.
At time $t=0$: (i) the first cluster band consists of the lowest $14$ bands, corresponding to two independent bosons in the lower sublattices;
(ii) the second cluster band ranges from the $15$-th band to the $43$-th band, corresponding to one boson in the lower sublattices and the other in the higher sublattices; and
(iii) the third cluster band ranges from the $44$-th band to the $57$-th band, corresponding to two independent bosons in the higher sublattices.

We then calculate the multi-particle Chern numbers for the two bound-state bands with Eq.~(3) in the main text~\cite{Takahiro2005}.
We obtain $+1$ and $-1$ for the highest two bands, which are consistent with the results obtained by the effective Hamiltonian.
The multi-particle Chern numbers for the first and third cluster bands are $1,-1$, respectively.
However, the multi-particle Chern numbers of the subbands from the second cluster band are not well defined, except for the $15$-th band with the Chern number $C_{15}=1$.

%%%%%%%%%%%%%%%
\begin{figure}[!htp]
\begin{center}
\includegraphics[width=1\columnwidth]{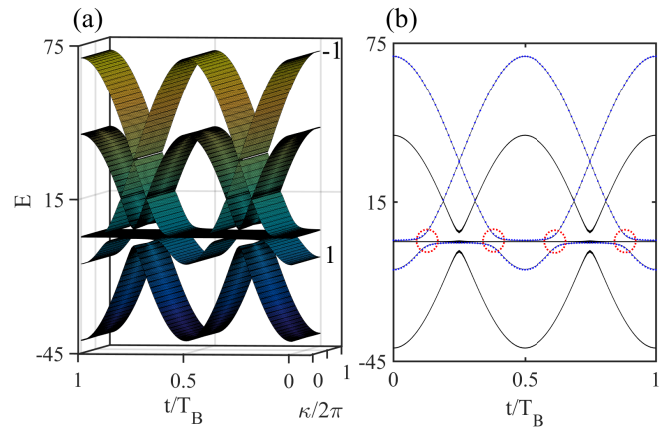}
\end{center}
\caption{Multi-particle Bloch band.
(a) The 3D view of Bloch band in the $\kappa-t$ Brillouin zone.
(b) The $t-E$ view of Bloch band. The parameters are chosen as $J=1,~\delta_0=0.8,~\Delta_0=20,~\varphi_0=0, ~U=30$ and the system size is 58. }
\label{Spectrum2}
\end{figure}
%%%%%%%%%%%%%%%

\subsubsection*{Breakdown of bound states} \label{MPBB2}

When $U \gg (J,~\delta)$ and $U$ is comparable to $2\Delta$, resonant tunneling happens between $\ket {2}_j$ and $\ket {1}_j\ket {1}_{j+1}$ or $\ket {1}_{j-1}\ket {1}_j$.
The coupling between the subspace $\mathcal U$ and $\mathcal V$ can not be neglected and the single quasi-particle (bound-state) approximation breaks down.

Increasing the double-well bias $2\Delta$ can significantly alter the multi-particle Bloch band, see Fig.~\ref{Spectrum2} (a) and (b) as the $t-E$ view of (a).
$\Delta_0$ is changed to $20$ and the other parameters keep the same as those for Fig.~\ref{Spectrum1}.
Compared to Fig.~\ref{Spectrum1}, the energy bands become much more complicated.
However, the well defined multi-particle Chern numbers do not change.
In particular, the $15$-th band with $C_{15}=1$ is separated from the second cluster band.
For comparison, we show the energy spectrum of two interacting bosons in a single double-well cell.
Such system is governed by the following Hamiltonian,
\begin{eqnarray}
 & {\hat H_{dw}} =   -(J + \delta_0 ) \hat a_{1}^\dag {\hat a_{2}}  + h.c. +\Delta_0 \cos(\omega t+\varphi_0) \left( {\hat n_2}-{\hat n_1}\right) \nonumber \\
 & +U/2\left[\hat n_1(\hat n_1-1)+\hat n_2(\hat n_2-1)\right].
 \label{HamDW}
\end{eqnarray}
The parameters are set as $J=1,~\delta_0=0.8,~\Delta_0=20,~\varphi_0=0$ and $U=30$, see the blue dots in Fig.~\ref{Spectrum2} (b).
The three energy spectra also partially cover multi-particle Bloch band.
At the energy anti-crossing marked by the red dashed circles, $2 |\Delta| = U$, the energy difference between $|1\rangle_1 |1\rangle_2$ and $|2\rangle_1$ or $|2\rangle_2$ is $0$.
Resonant tunnelings happen four times at the energy anti-crossing regime during one pumping cycle.
The bound states break down at the regime of resonant tunnelings.

%\bibliography{ResonantBound}

\begin{thebibliography}{44}
\bibitem {equalmomentum}%
\BibitemOpen {The dimension of the full Hilbert space is $\mathcal D={(N+qL-1)!/(N!(qL-1)!)}$ for our bosonic system. If $N$ and $L$ are not coprime numbers, the dimension for the block Hamiltonian $\hat H(\kappa_j)$ depends on the quasi-momentum $\kappa_j$. If $N$ and $L$ are coprime numbers, each block Hamiltonian $\hat H(\kappa_j)$ will have the same reduced dimension $\mathcal D_{s}=\mathcal D/L$} \BibitemShut {NoStop}%
\end{thebibliography}
%

\end{document}